\documentstyle[prl,aps,amssymb,twocolumn,floatfig,epsfig]{revtex}

\begin{document}

\draft
\title
{Bloch Electrons in a Magnetic Field: Why Does Chaos Send Electrons  the Hard Way?}
\author{R.~Ketzmerick, K.~Kruse, D.~Springsguth, and  T.~Geisel}

\address{
Max-Planck-Institut f\"ur Str\"omungsforschung und Institut f\"ur Nichtlineare Dynamik der
Universit\"at G\"ottingen, Bunsenstra{\ss}e 10, D-37073 G\"ottingen, Germany }

\maketitle
\begin{abstract}
We find that a 2D periodic potential with different modulation
amplitudes in $x-$ and $y-$direction and a perpendicular magnetic
field may lead to a transition to electron transport along the direction
of {\it stronger} modulation and to localization in the direction of
{\it weaker} modulation. In the experimentally accessible regime  
we relate this new quantum transport phenomenon to avoided band crossing due to classical chaos. 
\end{abstract}
\narrowtext

Theoretical studies of quantum chaos have recently included {\it quasiperiodic} systems   
concentrating so far on the kicked Harper model~\cite{kharper}. These studies 
have led to the discovery of interesting phenomena, e.g., 
surprising metal-insulator transitions~\cite{mit} whose origin are avoided band crossings induced 
by  classical
chaos~\cite{roland}. While the kicked Haper model is a numerically convenient toy model 
we here focus on a prominent example~\cite{peierls} from solid state physics which  
nowadays is experimentally accessible~\cite{exp}: an electron 
moving in a 2D periodic potential, a
so-called Bloch electron, subjected to a magnetic field. 
It is classically chaotic~\cite{theo} and it is quasiperiodic when the number of magnetic flux quanta per unit cell of the potential
is irrational~\cite{peierls}. 
 In this paper we present a new 2D quantum transport phenomenon: 
By changing the potential strength, keeping the ratio of the  
modulation strength in $x-$ and $y-$direction fixed, we    
find that transport exclusively along the direction of {\it weaker} potential modulation changes 
 to  transport exclusively along the direction of {\it stronger} modulation~(Fig.~1). 

Ballistic transport exclusively along the direction of {\it weaker} modulation is expected for the limiting cases of
either a very small or a very large potential strength compared to the magnetic field strength. 
This is based on the properties of the Harper model~\cite{harper,rauh,aubry,soko,barelli} which 
approximates Bloch electrons in a magnetic field in both limiting cases. 
For transitions to ballistic transport along the direction of {\it stronger} modulation we find two distinct 
 origins.  In the regime where potential and magnetic field are of comparable 
strength~\cite{regime} we show
that there are many transitions which we relate to avoided band crossings induced by the 
classical non-integrability. In the regime of large potential strength where the energy spectrum 
consists of separated bands there are additional transitions in some of these bands which 
can be explained in the tight-binding approximation and which are not related to the classical limit. 
Experimentally, the transitions of\noindent 
\begin{figure}[t]
\centering
\vspace*{-2.0cm}
\epsfig{figure=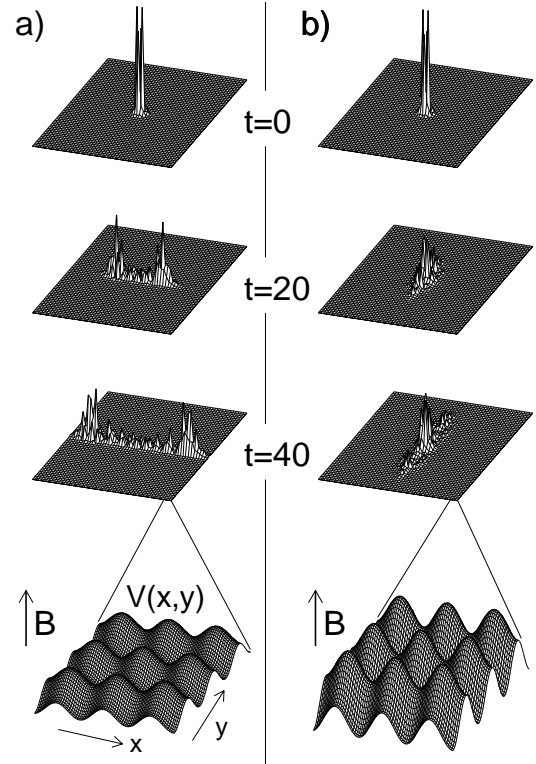,width=9.5cm}
\vspace*{-3.2cm}
\caption{
Time evolution for initially localized electron wave
packets~\protect\cite{f1} spreading in 2D periodic potentials with
different modulation amplitude in $x-$ and $y-$direction (bottom) and a perpendicular magnetic field. a) 
 For $K=5$ the wave packet spreads ballistically in the direction of {\it weaker}
modulation only, as expected for $K\rightarrow 0$ and $K\rightarrow\infty$. 
b) For  $K=10$ the wave
packet spreads ballistically in the direction of {\it stronger} modulation only. 
Shown are $55\times 55$ unit cells of
the potential,  times are in units of the
cyclotron period $2\pi/\omega_c$, 
$V_y/V_x=1.25,$ and $\Phi/\Phi_0=89/55$. 
}
\end{figure}
\noindent the transport direction in the first regime  should be 
observable with lateral surface superlattices on semiconductor heterojunctions.

The one-particle Hamilton operator for an electron with charge $-e$
and mass  $m$ in a magnetic field and in the simplest 2D periodic potential 
has the form
\begin{equation} \label{hamil}
H  =\frac{1}{2m}({\bf p}+e{\bf A})^{2}+V_{x}\cos(\frac{2\pi x}{a})+V_{y}\cos(\frac{2\pi y}{a})~.
\end{equation}
 The magnetic field ${\bf B}={\rm rot}{\bf A}$, where {\bf
A} is the vector potential, is taken to be homogeneous and
perpendicular to the $xy$-plane. The properties of the system depend
on the following three dimensionless parameters: 
the  potential strength 
\begin{equation}\label{KK}
K=(V_{x}+V_{y})\,4 \pi\, m\,a^{2}/h^2~,
\end{equation}
the ratio $V_{y}/V_{x}$, and the ratio of the magnetic flux
penetrating a unit cell of the potential divided by the magnetic flux
quantum,  $\Phi/\Phi_{0}=a^2B/(h/e)$, see e.g.\ Ref.\cite{dennis}.  
The ratio of the potential strength to magnetic field strength is given in
terms of these parameters by $2\,(V_{x}+V_{y})/(\hbar\omega_c)=K\,\Phi_0/\Phi$\,, with the cyclotron
frequency $\omega_c=e\,B/m$. 
In the following, we will vary
only $K$, while the other parameters will be kept fixed: $V_{y}/V_{x}=1.25$ and 
$\Phi/\Phi_{0}=(\sqrt{5}-1)/2$, the golden mean. 
Figure 1a shows the time evolution of an initially localized wave packet for $K=5$~\cite{f1}. 
 It spreads ballistically in the direction of weaker potential modulation ($x-$direction) and localizes in the
direction of stronger modulation ($y-$direction) as expected for $K\rightarrow 0$ and $K\rightarrow
\infty$. In contrast, for $K=10$ (Fig.~1b), 
localization occurs in the direction of {\it
weaker} modulation and one finds ballistic spreading in the direction of {\it stronger} modulation. 
 In other words, as a function of potential strength (keeping the ratio $V_y/V_x$ constant) 
 Fig.~1 shows that the system undergoes a metal-insulator transition in $x-$ 
direction and an insulator-metal transition in $y-$direction. 

\begin{figure}[t]
\centering
\epsfig{figure=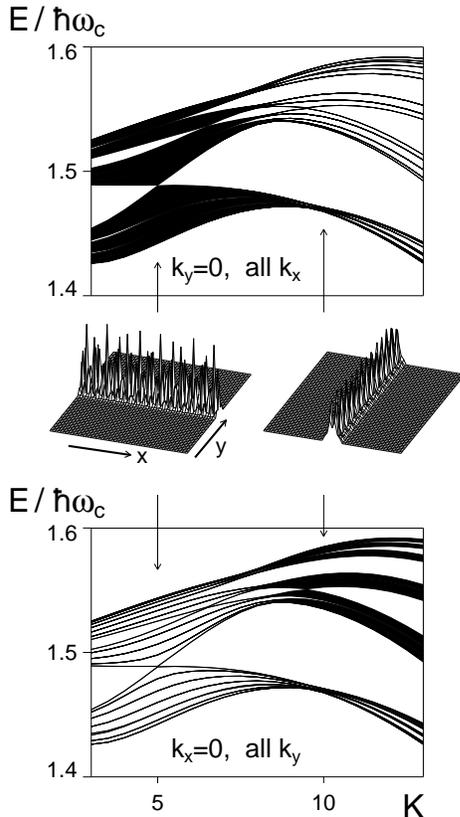,width=5.5cm}
\caption{
Subspectra for $k_y=0$ and all $k_x$~(top),  
 $k_x=0$ and all $k_y$~(bottom) vs. $K$, and typical eigenfunctions~\protect\cite{efct}
  for $K=5$ and $K=10$~($\Phi/\Phi_0=89/55$)\,.
}
\end{figure}
\newpage
We demonstrate that this change of the direction 
of transport is related to changes of spectrum
and eigenfunctions of operator (1). 
Numerically, one has to study rational approximants $\Phi/\Phi_0=q/p$. Then, operator (1) is
periodic in $x-$ and $y-$direction and its eigenenergies depend on the two phases $k_x$ and $k_y$
of the magnetic Brillouin zone. 
In Fig.~2 top (bottom), increasing $K$ from 5 to 10, one finds for the energy subspectrum with
only $k_x~(k_y)$  varied a transition from wide to extremely narrow (narrow to wide) minibands 
while eigenfunctions change
from extended to localized (localized to extended) in $x-$direction ($y-$direction)~\cite{efct}.  
These wide (extremely narrow) minibands correspond to an absolutely continuous (pure point) 
spectrum in the irrational case and we will abbreviate them in the following by 'bands' ('levels'). 
Figure 2 shows that eigenfunctions in $y-$direction and the subspectrum under variation of $k_y$
are dual to the corresponding behaviours in $x-$direction. The origin of this  duality
which we have found for all parameters studied seems to be related to the origin of the 
Aubry duality~\cite{aubry} of the Harper model and remains to be explored.
Using the duality we concentrate in
the following on the properties in $x-$direction. 
In Fig.~3 one can see the subspectrum under variation of $k_x$ on larger scales than in Fig.~2 top.
For $K=0$, one has Landau levels which for small $K$ broaden
linearly with $K$ forming Landau bands which have a fine structure  described by the Harper model (see
below). 
When increasing $K$, the spectrum becomes very complex: First of all, there are many transitions 
from 'bands' to 'levels' and vice versa. 
Secondly, one observes many avoided crossings of spectral regions, e.g., there is
one in the box corresponding to Fig.~2a. One notices that the spectral transitions happen in the
range of avoided crossings.  

We propose that these spectral transitions for Bloch electrons in a magnetic field are in fact due to 
the {\it avoided band crossings}:
These are analogous to avoided level crossings in classical chaotic, bounded quantum systems 
with a discrete spectrum and they occur in extended systems with a classically 
 non-integrable  Hamiltonian. 
In Ref.~\cite{roland} consequences of 
avoided band crossings were analyzed  by studying a three-band model,
in which each band was described by a tight-binding Hamiltonian, e.g., by a Harper model. 
In the range where the bands avoid crossing 
a perturbation calculation predicted that 
the parameters of their tight-binding  Hamiltonian effectively
change. In particular, when one of these tight-binding Hamiltonians is a quasiperiodic Harper model
and if the change is large enough this leads to  transitions from 'bands' to 'levels' or vice versa~\cite{roland}. 
In our case of Bloch electrons in a magnetic field 
such a perturbation analysis for isolated avoided crossings cannot be repeated as the 
energy spectrum is much too complex (Fig.~3). There is, however, a strong analogy: 
 There are spectral regions performing avoided crossings as this extended system 
is classically non-integrable~\cite{theo}. Secondly, before the cross-\\ \noindent 
\begin{figure}[t]
\centering
\epsfig{figure=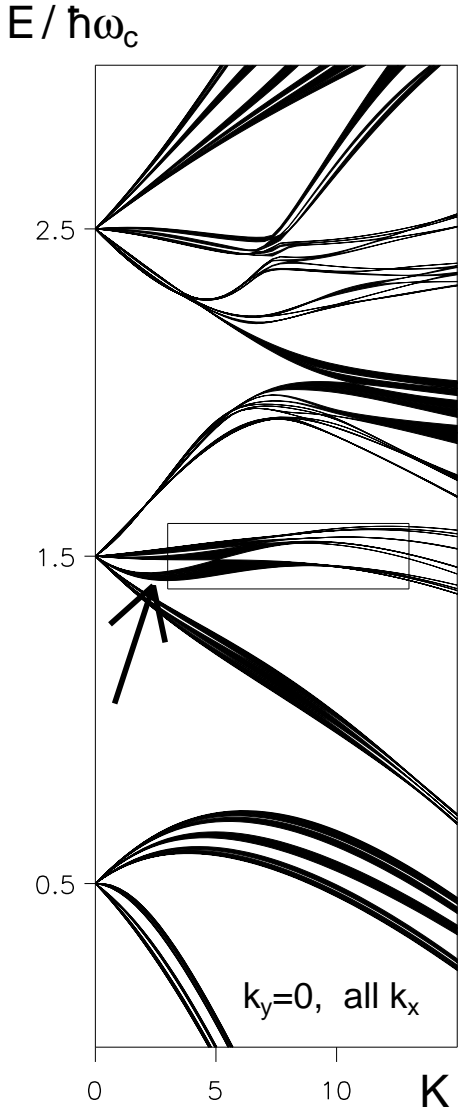,width=7.5cm}
\caption{
The subspectrum for $k_y=0$ and all $k_x$ vs. $K$~($\Phi/\Phi_0=34/21$)\,. 
Several avoided
crossings can be observed which are accompanied by transitions from energy 'bands' to
energy 'levels' or vice versa. The box indicates the region
shown in Fig.~2 and the arrow points to the spectral region  that
avoids crossing the one above. 
}
\end{figure}
\noindent ings (small $K$) these spectral 
regions may be described by Harper-like models~\cite{wilkinson}. Finally, we also find transitions 
from 'bands' to 'levels' or vice versa in the range of the avoided crossings.
 This analogy strongly suggests that also for Bloch electrons in a magnetic field 
 avoided band crossings due to classical chaos 
give rise to 
 the spectral transitions in Fig.~3, thereby 
inducing the change of
transport properties in $x-$direction in Fig.~1. 
Thus, finally, the change in transport is linked to the classical non-integrability 
of the Hamiltonian (1). 
It is important to stress that the classical dynamics shows  no corresponding change in 
transport  for these parameters. 
Instead, as usual in the field of quantum chaos,   
the classical dynamics enters via its non-integrability that here leads  to   
 the avoidance of band crossings.

There are exceptional spectral transitions for large $K$ which have a different origin: For 
strong potential, $K\gg \Phi/\Phi_0$, and for sufficiently low energy    
 the spectrum of Bloch electrons in a magnetic field consists of well
separated bands. In the tight-binding approximation  each of them has a fine structure 
described by Harper's equation~\cite{harper}  
\begin{equation}
\psi_{n+1}+\psi_{n-1}+2\,\lambda\,\cos(2\,\pi\,\sigma\,n\,+\nu)\,\psi_n=\tilde{E}\,\psi_n~, 
\end{equation}
 where $\lambda$ is the important parameter in the following. 
While  $\lambda>1$ corresponds to localization in $x-$direction,   
 $\lambda<1$ corresponds to spreading in $x-$direction, for typical irrational
$\Phi_0/\Phi$~\cite{soko}.  
These separated bands follow the 
 bands in the case {\it without} magnetic field in which the Hamiltonian separates 
in $x-$ and $y-$direction such that  each  band is the sum of an energy  band in $x-$ 
and one in $y-$direction. The 
$n_{\alpha}-$th band ($n_{\alpha}=0,1,2,\ldots)$ of the 1D potential
$V_{\alpha}\cos(2\,\pi\,\alpha/a)$  has a dispersion amplitude $\epsilon_{\alpha}(n_{\alpha})$, 
where $\alpha=x,\,y\,,$ 
that is determined by tunneling.   We will need that 
it increases with index $n_{\alpha}$ and that it decreases exponentially with potential 
strength $V_{\alpha}$. 
{\it With} magnetic field the transport properties of each  band, enumerated 
 by $(n_x,n_y)$, is determined in the tight-binding approximation by Harper's equation with 
$\lambda=\epsilon_y(n_y)/\epsilon_x(n_x)$. 
While for $K\ll \Phi/\Phi_0$, each Landau band has a fine structure again described by Harper's
equation [Eq.~(3)]~\cite{rauh} with a constant 
$\lambda=V_x/V_y$, here, $\lambda$ depends on the band indices $n_x,\,n_y$. There are two cases:
 For $n_y\leq n_x$ we find from the properties of $\epsilon_{\alpha}(n_{\alpha})$ that 
$\epsilon_y(n_y)<\epsilon_x(n_x)$ (for $V_y>V_x$) resulting in $\lambda<1$ as in the limit
$K\rightarrow 0$. 
 For $n_y> n_x$ we find that there exists a $K_c(n_x,n_y)$ where a transition occurs. For $K>K_c$
one still has $\epsilon_y(n_y)<\epsilon_x(n_x)\,$. In contrast, for $K<K_c$ the opposite is true
leading to $\lambda>1$ and thus to localization in $x-$direction. 
These transitions within a tight-binding band are a consequence of the $\lambda-$dependence 
of the Harper model and are not related to avoided band crossings or the classical dynamics. 
An example for such a transition is the tight-binding band $(0,1)$ for which 
$K_c\approx 1000$~\cite{trans}. In the subspectrum with $k_x$ varied it consists of 'bands' for 
$K>K_c$ while for $K<K_c$ we find 'levels' which  
appear, below the regime where the tight-binding approximation is valid,  
in Fig.~3 at $E/(\hbar\omega_c)\approx 1.5$, $K=15$.

In order to complete the picture on spectral transitions for Bloch electrons in a magnetic field  
let us mention that without magnetic field  for large $K$ pairs of  bands can make 
isolated crossings, e.g., 
the  bands with $(0,2)$ and $(1,1)$ at $K\approx 50$.  
In the presence of a magnetic field each of these crossings becomes an avoided crossing of Harper
models and 
may be well described by the models of Ref.~\cite{roland}
and may have many spectral transitions and corresponding changes in transport. Let us mention 
that  the properties of the tight-binding bands 
before and after an isolated crossing are  not affected.

Generalizing these results to arbitrary $V_y>V_x$ for Bloch electrons in a magnetic field~[Eq.~(1)] we
find that there are spectral transitions which have two distinct origins: 
{\bf i)} In the regime where the tight-binding description is valid the $K-$dependence of the band
widths $\epsilon_{\alpha}(n_{\alpha})$ and the properties of the Harper model predict spectral
transitions within  tight-binding bands with $n_y>n_x$ that are unrelated to the classical limit. 
In the subspectrum with $k_x$ varied 
 these transitions are from 'bands' to 'levels' only and not 
vice versa when decreasing $K$ from $\infty$. 
  In order to end up again with 'bands' for $K\rightarrow 0$ there have to be further 
transitions from 'levels' back to 'bands'. {\bf ii)} These transitions {\it and many more transitions in
both ways} appear in the 
regime where tight-binding
bands merge and form Landau bands (Fig.~3). We ascribe  them to avoided crossings 
of spectral regions which in turn are induced by classical chaos. 
Corresponding to the spectral transitions in the subspectrum when $k_x$ is varied we find dual 
transitions when $k_y$ is varied (Fig.~2). These transitions in both subspectra cause the 
observed change of transport from the direction of weaker to the one of stronger modulation
(Fig.~1). 
In general, transport in just one direction, as in Fig.~1, will occur whenever an initial wave packet
excites an energy range with eigenfunctions of just one type, namely either extended in $x-$ or in
$y-$direction. In the exceptional case that the small energy range of width kT around the Fermi energy 
contains both types of eigenfunctions one finds a superposition of transport in each direction, namely a
cross-like spreading. For other 2D periodic potentials, e.g., the potential of Eq.~(1) with different
lattice constants or rectangular antidot potentials, we find the same changes of the transport direction
in the regime of avoided band crossings~\cite{diss}.

These phenomena 
should be experimentally accessible using 
 lateral superlattices on semiconductor heterojunctions~\cite{exp} 
 by measuring the ratio $\rho_{yy}/\rho_{xx}=\sigma_{xx}/\sigma_{yy}$. 
Weak disorder as present in the experimental probes will destroy the transport phenomenon in the 
regime of the exponentially narrow tight-binding bands. Instead, in the regime where 
tight-binding
bands and Landau bands merge and avoided band crossing occur, 
 preliminary numerical studies show that the phenomenon should be observable.

\acknowledgments
This work was supported in part by the Deutsche Forschungsgemeinschaft.



\begin{thebibliography}{99}
\bibitem{f1}
 We construct the initial wave packet by taking $(x +
i\,y)\,\exp[-(x^2 +y^2)/(4\,l^2)]$, an eigenfunction of the second
Landau level for $K=0$, and projecting it on the subspace of
eigenfunctions for the given value of $K$ with energies in the
interval $[1.4\,\hbar\omega_c,1.6\,\hbar\omega_c]$. 
The figures were obtained by considering  the 4 lowest Landau bands~\cite{dennis}.
\bibitem{efct}
 The eigenfunctions shown in Fig.~2 are the 
linear superpositions of $p-$fold degenerate eigenfunctions excited by 
 the initially localized wave
packets of Fig.~1.
\bibitem{trans}
The transition at $K_c\approx 1000$ for $V_y/V_x=1.25$ is numerically not observable 
as the tight-binding band
width is exponentially small in $K$. For $V_y/V_x=2$, where $K_c\approx200$, we have observed such
a transition within the tight-binding band $(0,1)$.
\end{thebibliography}
\end{document}